\documentclass{aa}  
\usepackage{graphicx}
\begin{document}
\newcommand{\be}{\begin{equation}}
\newcommand{\ee}{\end{equation}}
\title{Particle acceleration by fluctuating electric 
fields at a magnetic field null point}

\author{Panagiota Petkaki 
\inst{1}
and
Alexander L. MacKinnon
\inst{2}}
\offprints{P. Petkaki}
\institute{\inst{1} Physical Sciences Division,
                  British Antarctic Survey, Cambridge, CB3 0ET, UK \\
                  \inst{2} DACE/Physics and Astronomy,
University of Glasgow, Glasgow, G12 8QQ, UK }

\date{Received date; accepted date}
\authorrunning{Petkaki \& MacKinnon}
\titlerunning{Time-Varying Electric Field}
 
  \abstract
   {Particle acceleration consequences from fluctuating electric 
fields superposed on an X-type magnetic field in collisionless solar plasma are studied.
Such a system is chosen to mimic generic features of dynamic reconnection, or the reconnective dissipation of a linear disturbance.}
   {Explore numerically the consequences for charged particle distributions of fluctuating electric 
fields superposed on an X-type magnetic field.}
   {Particle distributions are obtained by numerically integrating individual charged particle orbits 
when a time varying electric field is superimposed on a static
X-type neutral point. This configuration represents the effects of the 
passage of a generic MHD disturbance through such a system. 
Different frequencies of the electric field are
used, representing different possible types of wave. The electric field
reduces with increasing distance from the X-type neutral point as in linear dynamic magnetic reconnection.}
   {The resulting 
particle distributions have properties that depend on the amplitude and frequency of the electric field. In many 
cases a bimodal form is found. Depending on the timescale for variation of the electric field, electrons 
and ions may be accelerated to different degrees and often have energy distributions of different forms. Protons are accelerated to $\gamma$-ray producing energies and electrons to and above hard X-ray producing energies in timescales of 1 second. The acceleration mechanism is possibly important for solar flares and solar noise storms but is also applicable to all collisionless plasmas.}
   {}

\maketitle

\section{Introduction}

Release of stored magnetic energy via particle acceleration is a characteristic 
feature of astrophysical plasmas. In the particular case of the Sun, we see this 
manifested in the catastrophic events of flares, as well as in quieter phenomena 
like radio noise storms. Similar phenomena are observed in other late-type stars, 
and similar physics may be involved in understanding a wide variety of astrophysical 
objects (see e.g. Kuijpers \cite{Kuijpers93}, Hanasz and Lesch \cite{HanaszLesch}).
 
The special case of solar flares involves particular challenges to theory. 
A large fraction (several tens of percent) of the flare energy is 
manifested initially in the form of fast electrons (accelerated out of the background distribution
to $\sim $100 keV in about 1 second and to $\sim$ 100 MeV in a few seconds), which reveal 
their presence 
by producing bremsstrahlung X-rays (e.g. Miller \cite{Miller98}; MacKinnon \cite{MacKinnon06}). 
Protons are accelerated in flares to energies of several tens of
MeVs in a timescale of one second (Miller \cite{Miller98}, Aschwanden \cite{Aschwanden02}). 
Thus the acceleration of particles is an important 
part of the energy release process, rather than an energetically unimportant 
consequence of the flare. Moreover, radio signatures (Type I noise storms, 
Type III bursts away from flares) testify to particle acceleration at "quiet" times.

Magnetic reconnection is one of the primary candidate mechanisms for releasing
non-potential energy from magnetized plasmas (e.g. Priest and Forbes \cite{Priest01}).  
The electric field in
the current-carrying region also makes it a natural particle accelerator.
Collision-dominated sheets will involve the production of some runaway particles, but
almost by definition particle acceleration is not a primary consequence of such a
situation (e.g.  Smith \cite{Smith80}).  However, Martens (\cite{Martens88}) gave order-of-magnitude arguments
in favor of a collisionless current sheet as both the energy release mechanism and
the particle accelerator in flares.  Particle acceleration is energetically the primary
consequence of such a situation.  Collisionless reconnection thus assumes great potential
importance in understanding the flare process, particle acceleration, energy conversion and release in astrophysical plasmas
generally (Petkaki and MacKinnon \cite{Petkaki97}; Heerikhuisen et al. \cite{Heerikhuisen02}; Hamilton et al. \cite{Hamilton03}; Turkmani et al. \cite{Turkmami06}; Wood and Neukirch \cite{Wood05}; Vainchtein et al. \cite{Vainchtein05};  McClements et al. \cite{McClements06}).

Here we present test particle calculations designed to illuminate the consequences for
particle acceleration of dynamic reconnection.  We have in mind
particularly the picture of Craig and McClymont (\cite{Craig91}, \cite{Craig93}), in which a linear
disturbance passes through a magnetic configuration containing an X-type neutral point.
The disturbance travels non-dissipatively with the local Alfv\'{e}n speed until it
approaches the dissipation region surrounding the neutral point, where the resistive
diffusion term in the induction equation becomes important.  The wave damps
resistively in a few system transit times, with consequences (heating or particle acceleration)
determined by
the physical nature of the resistivity. Several assumptions were made in Craig
and McClymont's original discussion (linear disturbance, cold plasma limit, 2-D, Ohm's law 
including only a scalar resistivity)
but this essential picture still affords a qualitative guide in more complex situations 
(e.g. McClymont and Craig \cite{McClymont96}; Senanayake and Craig \cite{SenCraig06}).

In Petkaki and MacKinnon (\cite{Petkaki97}), we examined the
behavior of protons in the presence of electric and magnetic
fields obtained from the Craig and McClymont (\cite{Craig91}) analysis. 
Here we carry out a complementary exercise, studying test particle 
evolution in the presence of simple fields chosen to mimic generic features of dynamic reconnection. Our aim is to comment on
particle acceleration consequences, in a parametric
way that does not depend on a particular set of simplifying physical assumptions 
or boundary conditions. Time-dependence of the electric field is the essential ingredient
reflecting the dynamic character of the reconnection. We present examples of distributions
resulting from a time-independent electric field for comparison and highlight
distinct features of the distributions resulting from dynamic situations.

We use the Craig and McClymont (\cite{Craig91}) linear solution as a qualitative guide for the spatial 
and temporal form of the electric field. Our adopted field also resembles a linear situation in 
displaying a time dependence that does not change (i.e. does not develop multiple frequencies, saturate, etc.).
 Basing our calculations on this linear 
picture makes it unlikely that they will provide a complete description of what happens in 
a flare, although they offer useful insight. They may however be particularly relevant to non-flaring particle 
acceleration, e.g. in solar noise storms, or as part of the explanation of 'quiescent'
radio emission seen from RS CVn binaries (Kuijpers \& van der Hulst \cite{Kuijpers85}).

Since we aim to emulate a linear situation we may pick our test particles
from an isotropic, homogeneous distribution representing the background. This is in contrast to particle studies of
nonlinear reconnection, where consistency demands consideration
of the motion of particles into the dissipation region. In most studies
particles are injected in two opposite quadrupoles of the X-point and
they subsequently are driven, by the ${\b E} \times {\b B}$ drift due
to an imposed constant electric field, to cross the nonadiabatic
region or miss it depending on their initial conditions (see e.g. 
Burkhart et al. \cite{Burkhart90}).

Many previous studies of test particle evolution in steady reconnection exist. 
Here we mention particularly the work of Martin (\cite{Martin86}), which demonstrates that the orbits 
of such test particles are chaotic, and of Burkhart et al. (\cite{Burkhart90}, 
\cite{Burkhart91}) who iterated from the test particle calculations to construct 
a self-consistent description of the diffusion region. Recent work  
studies regular and chaotic dynamics in 3-D reconnecting current sheets
(Efthymiopoulos et al. \cite{Efthym05}; Gontikakis et al. \cite{Gonti06}) or studies
particle orbits in the presence of 3-D magnetic nulls (Heerikhuisen et al. \cite{Heerikhuisen02}; 
Dalla and Browning \cite{dalla05}). Particularly relevant here is the exploratory, analytical study
of Litvinenko (\cite{Litvinenko03}) which looks at charged particle 
orbits in an oscillating electric field in a magnetic field containing a neutral line. 

The next section gives details of the specific model we adopt in order to study particle acceleration
in time-dependent reconnection, while Section~\ref{sec:secres} describes our results for particle distributions.
Section~\ref{sec:discon} discusses some possible implications of our results.

\section{Model for Particle Acceleration in fluctuating electric fields} 
\label{sec:model}

We are going to study the evolution of test particles 
in the presence of electromagnetic fields chosen to mimic generic features of dynamic reconnection. Time-dependence of the electric field 
reflects the dynamic character of the reconnection. 

\subsection{Equations of motion}

We solve numerically the relativistic equations of motions of test particles 
(particles are expected to acquire relativistic velocities) in electromagnetic fields and
in the observer's reference frame: 
\be
{{d{\b r}} \over {dt}} = {{\b p} \over {m \gamma}} \label{eq:lor1}
\ee
\be
{{d{\b p}} \over {dt}} = {q}{({{\b E} + {1 \over c}{( \b u \times \b B)}})} \label{eq:lor2}
\ee
where $\gamma = (1-(u/c)^2)^{-1/2}$, 
$ {\b u} = {\b p}/m \gamma $.
 
To model the reconnection magnetic field, we adopt an idealized 2-D magnetic field  containing an X-type neutral point: 
\be
{\b B} = {{B_0} \over D}({y {\hat x} + x {\hat y}}).
\ee
The current density vanishes for this field configuration. The field 
lines are the solutions of  
${dx \over {dy}}= {y \over x} $ which are hyperbolae $ y^2-x^2= const$.  
The X-line (neutral line) lies along the z-axis. The field strength depends on position
thus:  
\be{\vert {\b B} \vert}= B_0{ r \over D}\ee 
where  $x^2 + y^2 = r^2.$
Note that this configuration has no natural scale length.  Requiring
the field to have a value of $10^2$ gauss at a typical active region
distance of $10^9$ cm from the neutral point, fixes only $B_0/D =
10^{-7}$    gauss ${\rm cm^{-1}}.$ We are free to use other
considerations to fix one of $B_0$ and $D$ independently, as we do
below in introducing dimensionless variables.
An electric field is imposed in the $z$ direction,
with spatial and temporal form chosen to mimic qualitative features of dynamic
reconnection (see Sec. \ref{sec:Dyna}).

We normalize distances to $D_n$ and times to the gyroperiod at $r=D_n$. 
We denote the resulting timescales by $\tau_p$ and $\tau_e$ for the cases of electrons and protons
respectively. As noted above, $D_n$ is as yet undetermined. It turns out to 
be convenient
in this relativistic calculation to choose $D_n$ such that velocities are 
normalized to the speed of light. This has the consequence that $D_n$ takes 
different values $D_e = c\surd(m_e D/e B_0 )$ and $D_p= c\surd(m_p D/e B_0 )$ for electrons and protons respectively (Petkaki and MacKinnon, 
\cite{Petkaki94}, Petkaki, \cite{Petkaki96}), 
such that
\be
D_p = ({\frac{m_p}{m_e}})^{1 \over 2} D_e.
\ee
Specifically, with $B_o/D = 10^{-7}$, we find $D_e=1.3 \times 10^5$ cm and
$D_p=5.6 \times 10^6$  cm. 
With our choices of ${\b E}$ and ${\b B}$  the Lorentz equations (\ref{eq:lor1}) and (\ref{eq:lor2})
become in three dimensions and in dimensionless units:

\begin{eqnarray}
{{d{ \bar{x}}} \over {d \bar{t}}} & = & {\bar{u}_x} = {{\bar{p}_x} \over \gamma} \nonumber \\
{{d{ \bar{y}}} \over {d \bar{t}}} & = & {\bar{u}_y} = {{\bar{p}_y} \over \gamma} \nonumber \\
{{d{ \bar{z}}} \over {d \bar{t}}} & = & {\bar{u}_z} = {{\bar{p}_z} \over \gamma} \nonumber \\
{{d{\bar{p}_x}} \over {d\bar{t}}} & = & - \epsilon  \bar{x} \bar{u}_z \nonumber \\
{{d{\bar{p}_y}} \over {d\bar{t}}} & = & \epsilon  \bar{y} \bar{u}_z \nonumber \\
{{d{\bar{p}_z}} \over {d\bar{t}}} & = & \bar{E} + \epsilon (\bar{x} {\bar{u}_x} +
\bar{y} {\bar{u}_y})
\label{eq:mot-equ}
\end{eqnarray} 
where $\gamma = {{(1+{{{p}_x}}^2+{{{p}_y}}^2+{{{p}_z}}^2})}^{{\frac{1}{2}}}$ 
and 
$\epsilon=+1$ for protons,  $\epsilon=-1$ for electrons. 
$E({\b r},t)$ is the true value of the electric field and $\bar{E}= E (D/B_o) D_i$ is
the dimensionless electric field, with subscript $i$ taking the values $e$ for electrons
or $p$ for protons. 
Energies are now normalised to the particle rest mass energy so that kinetic energy 
in dimensionless units is just $K_{kin}=\gamma -1$. Equations (\ref{eq:mot-equ})  
with appropriate initial
conditions and a specific form for $\bar{E}$ describe the motion of a particle.

\subsection{Electric Field} 
\label{sec:Dyna}

\begin{figure}
\centering
   \includegraphics[width=0.45\textwidth]{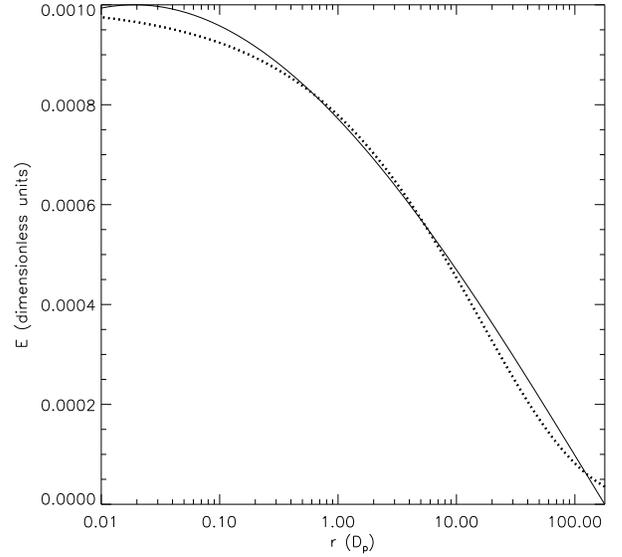}
\caption{Electric field (solid line) calculated from the Craig and McClymont (\cite{Craig91}) solution for inertial resistivity $\eta=3.1724 \times  10^{-11}$ and for the fundamental mode of azimuthal symmetry (n=0) at t=0. Approximate electric field described in Equation \ref{eq:Heav} is shown in dotted line at t=0.}
\label{fig:Electric}
\end{figure}

Craig and McClymont (\cite{Craig91}) guide us in adopting a functional form of electric field which 
allows us to investigate consequences of time-dependence in a parametric way. Their resistively damping, linear disturbance involves a regularly 
oscillating electric field whose amplitude is greatest in the region 
near the neutral point where the resistive term of the induction 
equation becomes important. Far from this region the disturbance is 
Alfv{'}enic in character and dominated by the boundary conditions, so 
that the electric field amplitude always maximises in the central, 
diffusion region (see also Petkaki and MacKinnon \cite{Petkaki97}, 
Section 3). Thus we adopt the following form
for the electric field ${\b E}$:
\be
{\b E} = E_0 \sin (\omega t) {\hat z} f(x,y)
\ee
where $f(x,y)$ describes the spatial variation of ${\b E}$. We take
\be
f(x,y)=exp(- \alpha_i \surd( |r|))
\label{eq:Heav}
\ee
where, $\alpha_p = 2.5 \times 10^{-1}$, $\alpha_e = 3.776 \times 10^{-2}$, $|r|=\surd(x^2+y^2)$. 
Consistent with our 
concentration on particle acceleration near the neutral point and with {\it in situ} measurements 
in Earth's magnetosphere ({\O}ieroset et al., \cite{oieroset01}), we expect that resistivity 
will be primarily inertial in character (Speiser \cite{Speiser70}). As shown in Figure~\ref{fig:Electric}, this
form approximates the radial form of the electric field calculated from the Craig and McClymont (\cite{Craig91}) 
solution, for an (inertial) resistivity $\eta$ estimated assuming the dominant contribution from 1 keV protons
(see Petkaki and MacKinnon \cite{Petkaki97}, 
Section 3; Speiser, \cite{Speiser70}). The form of the electric field is shown in Figure~\ref{fig:Electric} at $t=0$.
The exact Craig and McClymont solution develops more complex spatial structure but the spatial form of
Eq.~(\ref{eq:Heav}) thus embodies a dissipation region on the appropriate length scale.

The frequency of oscillation of the electric field 
which we denote by $ \omega$ is a free parameter. Each simulation uses one value of $\omega$.
We take values of $ \omega$ such that
$1/1000  < \omega < 10000$, corresponding to a broadband wave spectrum which
may propagate in such a system (see Petkaki and Mackinnon \cite{Petkaki97}).

To compare with a simple, unvarying state, we also calculate the energy distributions that result from  a constant in time imposed electric field
\be
{\b E} = E_0  {\hat z} f(x,y)
\label{eq:const}
\ee
where $f(x,y)$ is defined in Equation \ref{eq:Heav}. With its nonzero curl, this assumed form
of $E$ cannot represent a steady state reconnection. We employ it primarily to provide a simple, unvarying state for comparison with results in the 
time-dependent situation.

Key to understanding particle behaviour near the neutral point is the `adiabaticity' radius $r_{ad}$, the distance from
the neutral point at which the Larmor radius equals the magnetic field
scale length. For $r > r_{ad}$, particles move adiabatically. In other words, if the distance of the particle from the 
neutral point is of the order of its Larmor radius, then the particle is non-adiabatic. The
`adiabaticity' radius depends on the particle mass and velocity
perpendicular to the magnetic field, ${u_\perp}$, and is given  by
\be
 r_{ad}= ({{ m c D {u_\perp}} \over {e {B_o}}})^{1 \over 2} .\label{eq:radiab}
\ee
For electrons and
protons of the same energy, the electron gyroradius is
$({\frac{m_e}{m_p}})^{1 \over 2}$ smaller than the proton gyroradius.

\subsection{Numerical Method}\label{sec:Num}

Due to the complexity of the orbits, their calculation cannot be done 
analytically. For integrating the ordinary differential equations
(ODEs) describing the motion of the particles, we use the
Bulirsch-Stoer method (Press et al. \cite{Press96}). 
For a single particle with the same initial conditions, the orbit changes 
if the accuracy required of the integration routine is varied, and when the
particle crosses the neutral point area (for general properties of
X-type neutral point orbits see e.g. Martin \cite{Martin86}). 
The statistical properties of the {\it distribution}
of test particles, which are of primary interest here, are unaffected 
by changes in the accuracy required of the integration routine. In the absence
of an electric field, the routine conserves particle energy to one part in $ 10^{-5}$.

We start the integration of particle orbits at t=0 and with the particles 
positioned randomly in a box with the following size
\begin{eqnarray}
       -1.0 \le & x0 & \le 1.0  \\
       -1.0 \le & y0 & \le 1.0  \\
                & z0 & = 0.0   
\end{eqnarray}
in electron or proton units depending on the species. We integrate the particle orbits 
up to 230400 timesteps ($\tau_e$) for
electrons and  5360 ($\tau_p$) for protons. With $B_0/D = 10^{-7}$ and our
form of dimensionless units these times correspond to 1 second for electrons
and protons. 
The initial velocities of the particles are picked randomly from a
Maxwellian distribution of temperature $5 \times 10^6 K$ ($\sim 431$ eV), a typical
coronal value. 
We consider only small values for $\bar{E}_0$, consistent with the
passage of a disturbance in the linear regime (Craig and
McClymont, \cite{Craig91}). Values of 0.0001, 0.001 and 0.01 are used in the actual
calculation. The value 0.001 corresponds to electric field $=5.88
\times 10^{-4}$ statvolt/cm. These are moderate values for 
electric fields present in the solar atmosphere (see Foukal et al.,
\cite{Foukal86}).

\subsection{Particle Orbits}\label{sec:Orbits}

\begin{figure}
\centering
   \includegraphics[width=0.45\textwidth]{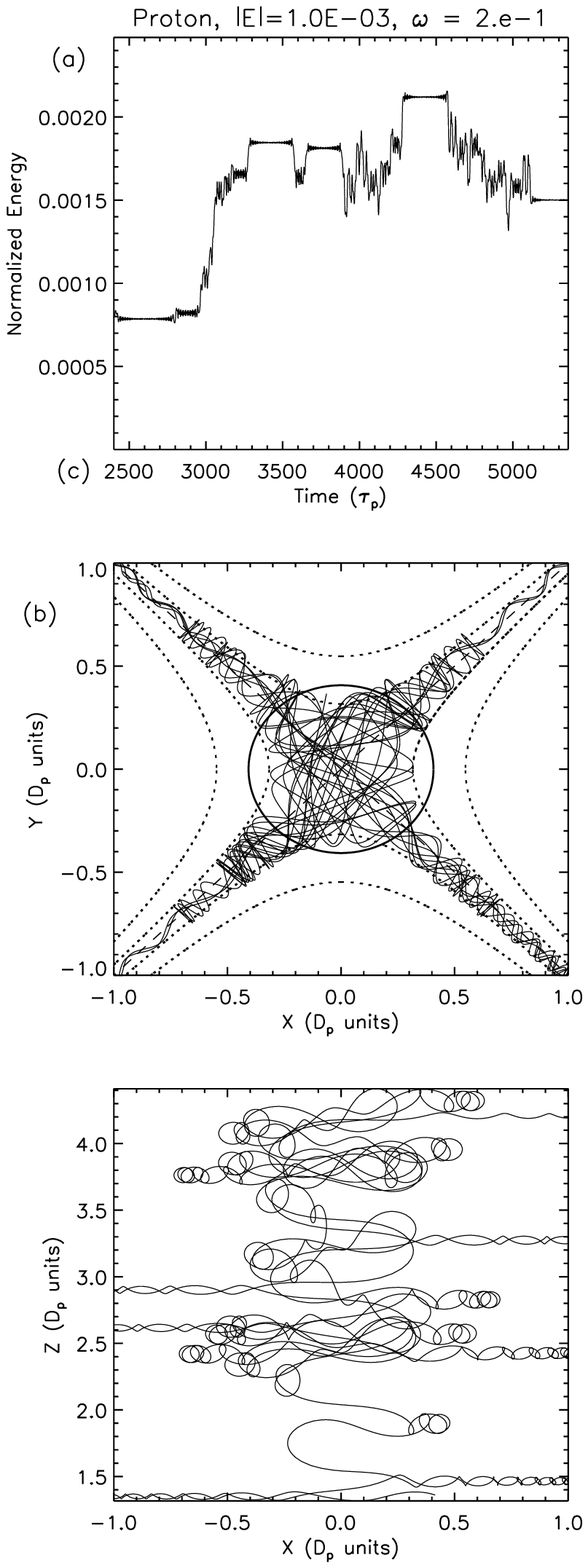}
\caption{Proton orbit in time-varying electric field of $\omega = 0.2$. (a) Energy as a function of time 
(b) Projection in the X-Y plane.  (c) Projection in the X-Z plane.}
\label{fig:POrb-ptwo}
\end{figure}

We are going to examine a typical proton orbit which is shown in
Figure \ref{fig:POrb-ptwo}. The amplitude of the time-varying electric 
field is $\bar{E}_0 = 0.001$ and the frequency is $\omega = 0.2$. 
In Fig. \ref{fig:POrb-ptwo}a we plot the  (dimensionless) energy of the proton
as a function of time for the interval 2400-5360 $\tau_p$. 
In Fig. \ref{fig:POrb-ptwo}b is shown the projection of the same
orbit on the X-Y plane and in Fig.~\ref{fig:POrb-ptwo}c the projection of the same
orbit on the X-Z plane. Our model assumes a system scale of $\sim 178 D_p$, corresponding to
a typical active region scale of $10^9$ cm. 
In Fig. \ref{fig:POrb-ptwo}b we zoom in close to the neutral point to observe the particle orbit in detail and 
we look in an area of $-1.0 D_p < \bar{x} < 1.0 D_p$ and $-1.0 D_p < \bar{y} < 1.0 D_p$. 
In the same figure superposed in dotted line are some of the magnetic field lines 
showing the structure of the X-type magnetic neutral point. All field lines tend to the
separatrices (shown in dashed lines) as the distance from the  neutral point becomes very
large. 

The thick solid circle has radius $d = 10 \space r_{ad} $ for a thermal proton. Inside this region the gyroradius (Larmor
radius) of most particles is not well defined since the particle is not bound to one magnetic field line 
and meandering motion is observed. The electric field accelerates or 
decelerates the proton causing further changes in the particle gyroradius and energy.
This behavior resembles a stochastic-type acceleration. Stochasticity is introduced by the phase of the electric field  
and the phase of the particle orbit and is sustained because of the form of the magnetic field 
(e.g. Martin \cite{Martin86}). Outside the magnetic neutral point area the particle is moving 
along a particular magnetic field line.  The gyroradius and the parallel velocity decrease as the particle
moves away from the neutral point. The particle mirrors and recrosses
the non-adiabatic region and the process is repeated until the end of the integration time or until the
particle escapes the outer boundary of the system ($x , y \ge 178 D_p$). 

Looking back at Fig.~\ref{fig:POrb-ptwo}a we see the variation of the particle energy as a function of time.
The intervals of energy conservation correspond to the times the particle is away from the 
nonadiabatic region, and in regions where the electric field is decreasing. Changes in the particle energy take place during the 
crossing of the non-adiabatic region.

The orbits of electrons show similar properties to that of the protons.
Additionally from the way
we pick our initial conditions the electrons start closer to the
neutral point (see Sect. \ref{sec:Dyna}). 
But since the form of the electric field for both species is calculated assuming proton inertial resistivity, 
electrons see an area much larger than their adiabaticity
radius where the electric field is close to its maximum value. Consequently some of the electrons
start their motion outside their adiabaticity radius and their motion is immediately
adiabatic. In this case the presence of the electric field does not
increase their energy except if they eventually cross the non-adiabatic region. 

The amount of
acceleration that particles get depends on the time they spend close to
the neutral point, on the phase of the orbit and on the frequency of the electric field. We define crossing time as the time 
the particle needs to cross
the non-adiabatic region (Sec.~\ref{sec:Dyna}) and is given to order of
magnitude by
\be
t_{cr} \sim \frac{2 r_{ad}}{u_{x,y}}  \label{eq:cross}
\ee
where $u_{x,y}$ is the velocity projection in the x-y plane. So,
\be
t_{cr} \sim {(\frac{2.828 c D}{e B_o})}^{1/2} \frac{m_i^{3/4}}{E^{1/4}}.
\ee
It turns
out that particles with the same energy satisfy
\be t_{{cr}_p} = t_{{cr}_e}
({\frac{m_e}{m_p}})^{3 \over 4}=280 t_{{cr}_e}
\ee
where $t_{{cr}_p}$ is
the proton crossing time and $t_{{cr}_e}$ is the electron crossing
time.  One would expect that in order to
get particles effectively accelerated (or decelerated since the sign of the
electric field is not constant) we need $ t_E \gg t_{cr}$ where $t_E=1/\omega$
is the period of fluctuation of the electric field. Thus we see potential
for differences between electron and ion acceleration.

\section{Energy Distributions of Accelerated Electrons and Protons}
\label{sec:secres}

We calculate the kinetic energy of each particle up to maximum of a
1 second real time along with final positions and velocities.
Particles in our calculation spend a 
relatively short time close to
the neutral point but they get trapped in the magnetic configuration
and re-cross the neutral point a number of times (Sect.~\ref{sec:Orbits}). 
Particles encounter the non-adiabatic region, the process resulting in a Fermi-type acceleration. 
A similar phenomenon has been
noted for a multiple neutral point configuration by Kliem (\cite{Kliem94}), 
in the behavior of protons in the presence of an MHD disturbance by Petkaki and Mackinnon (\cite{Petkaki97}), and was
explored analytically for time-varying electric field by Litvinenko (\cite{Litvinenko03}).

\begin{figure*}
\centering
   \includegraphics[width=17cm]{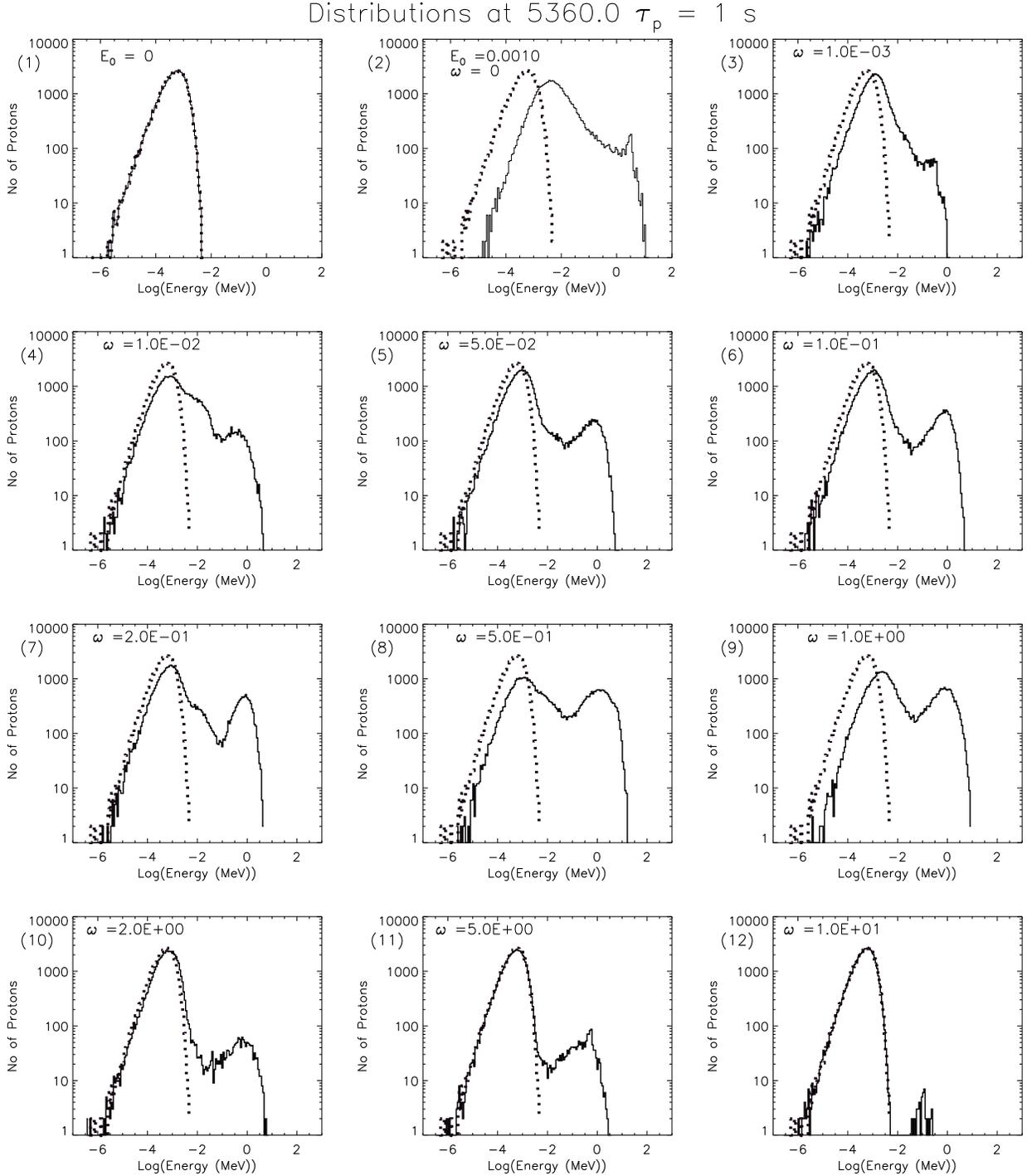}
\caption{Proton distributions for different frequencies of the 
electric field. The magnitude of the electric field
is $\bar{E}_0=0.001$. The total integration time is 5360.
\label{fig:Pro-energy}}
\end{figure*}

We sample the electric field frequency range 
$0.001  < \omega < 100$ for protons and $0.005  < \omega < 1000$ for electrons.
In Fig.~\ref{fig:Pro-energy} we plot histograms of the logarithm of the initial and 
final energy distributions of protons, for magnitude of the electric field $\bar{E}_0=0.001$ 
and total number of timesteps 5360. Each distribution is generated using 50000 test protons. 
We also calculated the distributions resulting when the
magnitude of the electric field is $\bar{E}_0=0.01$ and $\bar{E}_0=0.0001$.
The initial Maxwellian distribution is shown in dotted lines in each panel. 
Panel (1) shows the distributions for electric field magnitude $\bar{E}_0=0$. 
We observe no change in the form of the distribution since no acceleration is taking place 
(see also Sect.~\ref{sec:Num}). Panel (2) shows the energy distributions for constant electric field ($ \omega =0$) to provide a comparison with steady state magnetic reconnection. The final energy distribution for $ \omega =0$
has two distinct peaks, one at the initial Maxwellian distribution and a beamlike distribution close to 
$K_{kin}=10$ MeV. The energy distributions for constant electric field are not power laws as found elsewhere (see e.g.
Bulanov and Sasarov \cite{Bulanov76}; Bruhwiler and Zweibel \cite{Bruhwiler92}). In
those former calculations the particles crossed the neutral point 
only once, whereas particles recross the neutral point numerous times in our model.
The frequency of the time-varying electric field increases progressively from 
$\omega =0.001$ in panel (3) to $\omega =10$ in panel (12). For $\omega =0.001$ a small beamlike structure appears at $K_{kin}=0.3$ MeV.
The final energy distributions are bi-modal from $\omega =0.01$ to $ \omega =10$. For higher frequencies the proton energy distributions
do not show significant energy changes.

\begin{figure}
\centering
   \includegraphics[width=0.45\textwidth]{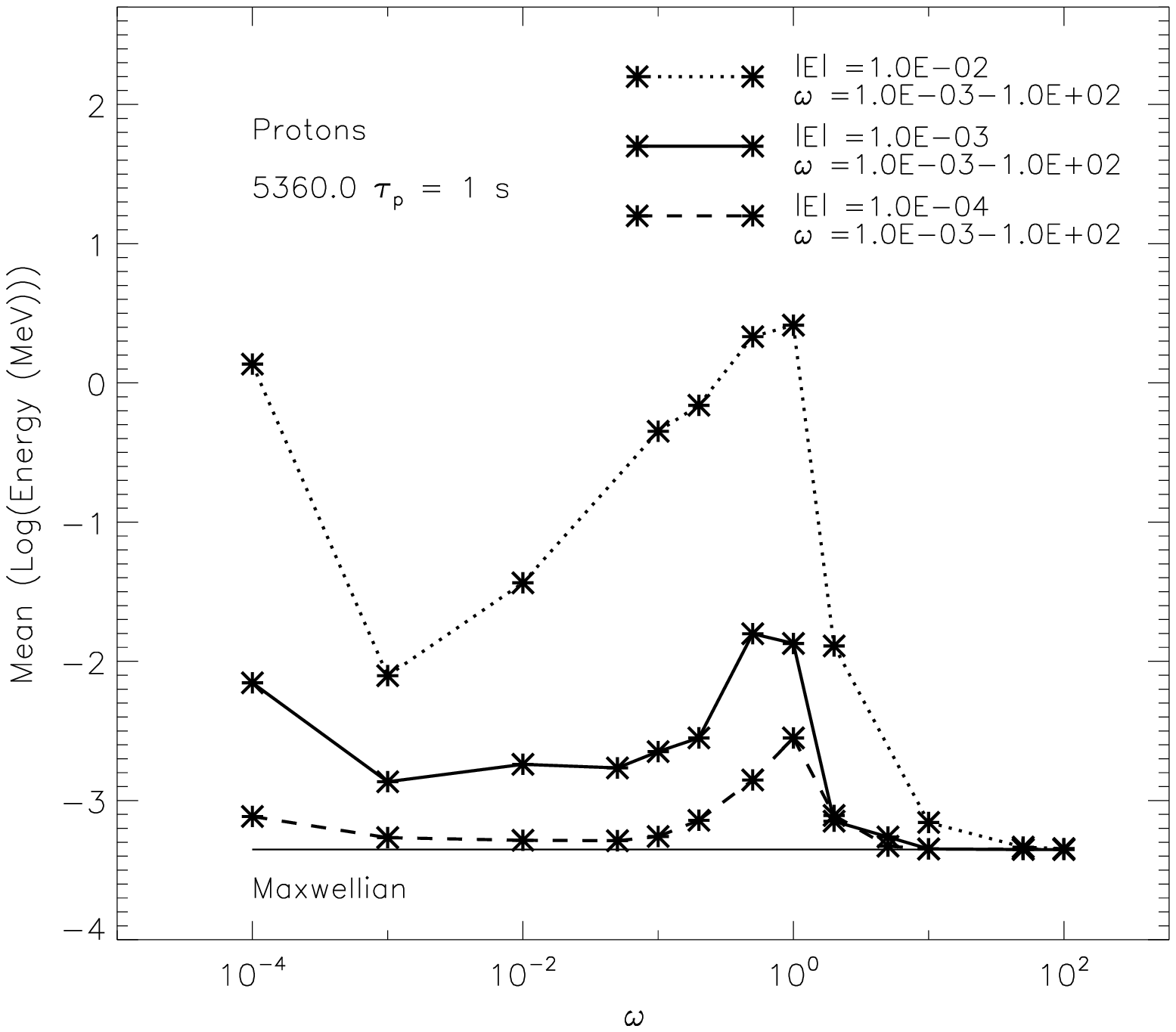}
\caption{Mean Energy of proton distributions for three amplitudes of the 
electric field ($\bar{E}_0 = 0.0001, 0.001, 0.01$) and for range of frequencies 0.001 to 100.0. 
The mean energy of the initial Maxwellian distribution is shown as a straight full line.
The constant electric
field case is represented by $\omega = 0.0001$. }
\label{fig:MeanPro-x234}
\end{figure}

\begin{figure}
\centering
   \includegraphics[width=0.45\textwidth]{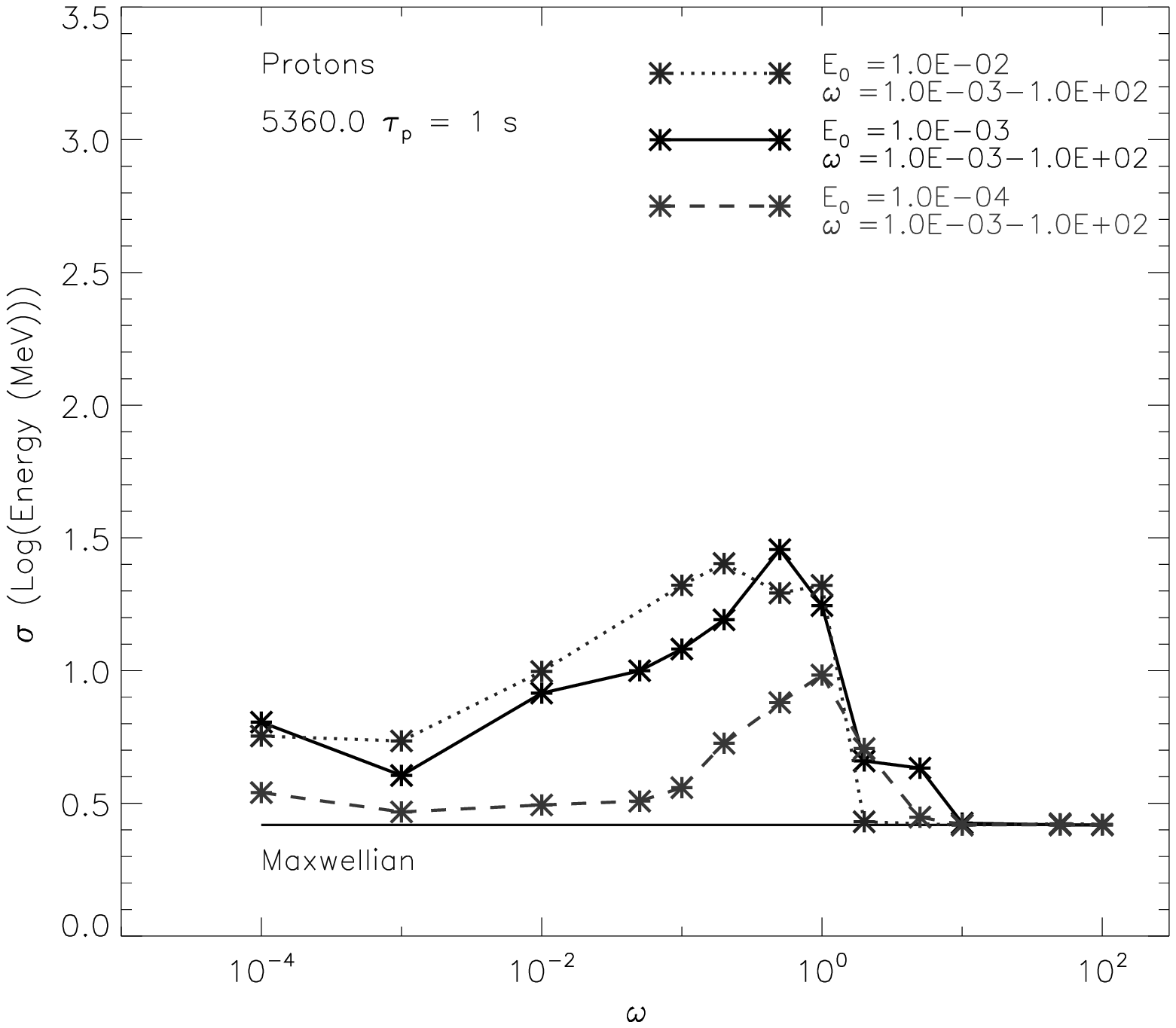}
\caption{Standard deviation of proton distributions for three amplitudes of the 
electric field ($\bar{E}_0 = 0.01, 0.001, 0.0001$) and for range of frequencies 0.001 to 100.0. 
The standard deviation of the initial Maxwellian distribution is shown as a straight full line. 
The constant electric
field case is represented by $\omega = 0.0001$.}
\label{fig:StdPro-x234}
\end{figure}

In Fig.~\ref{fig:MeanPro-x234} we plot the mean of the logarithm of the initial and
final proton energy distributions versus the frequency of the electric field and
for three amplitudes of the 
electric field $\bar{E}_0 = 0.0001$ (dashed star line), $\bar{E}_0 = 0.001$ (solid star line), 
and $\bar{E}_0 = 0.01$ (dotted star line). The mean energy for the constant
electric field is represented on this plot by $\omega= 10^{-4}$. The same representation
is used in Fig.~\ref{fig:StdPro-x234} where
we plot the standard deviation of the logarithm of the initial and
final proton energy distributions versus $\omega$ 
for the same three amplitudes of the electric field as in Fig.~\ref{fig:MeanPro-x234}. We use the
mean value of the logarithm of the energy to better represent the changes in highly non-thermal 
distributions.

The mean energy increases monotonically with $\bar{E}_0$ for the constant electric field case
and for all frequencies of the electric field except for the highest frequency
used in our model. The highest mean energy is achieved for constant electric field for all values
of the electric field.
For $\bar{E}_0 = 0.0001$ the highest energy gain for the time-varying electric field 
is achieved when $0.2 < \omega < 2.0$, indicating a resonant acceleration process.
For $\bar{E}_0 = 0.001$ a peak in the mean energy is also present when $0.2 < \omega < 2.0$.
For $\bar{E}_0 = 0.001$ protons gain most energy from the low $\omega$ electric field
(see Fig.~\ref{fig:MeanPro-x234}). When $\bar{E}_0 = 0.01$ protons a peak in the mean energy is present when $0.01 < \omega < 2.0$.
For frequency $\omega= 10$ and greater
the energy distribution does not
change significantly for all values of $\bar{E}_0$.

\begin{figure*}
\centering
   \includegraphics[width=17cm]{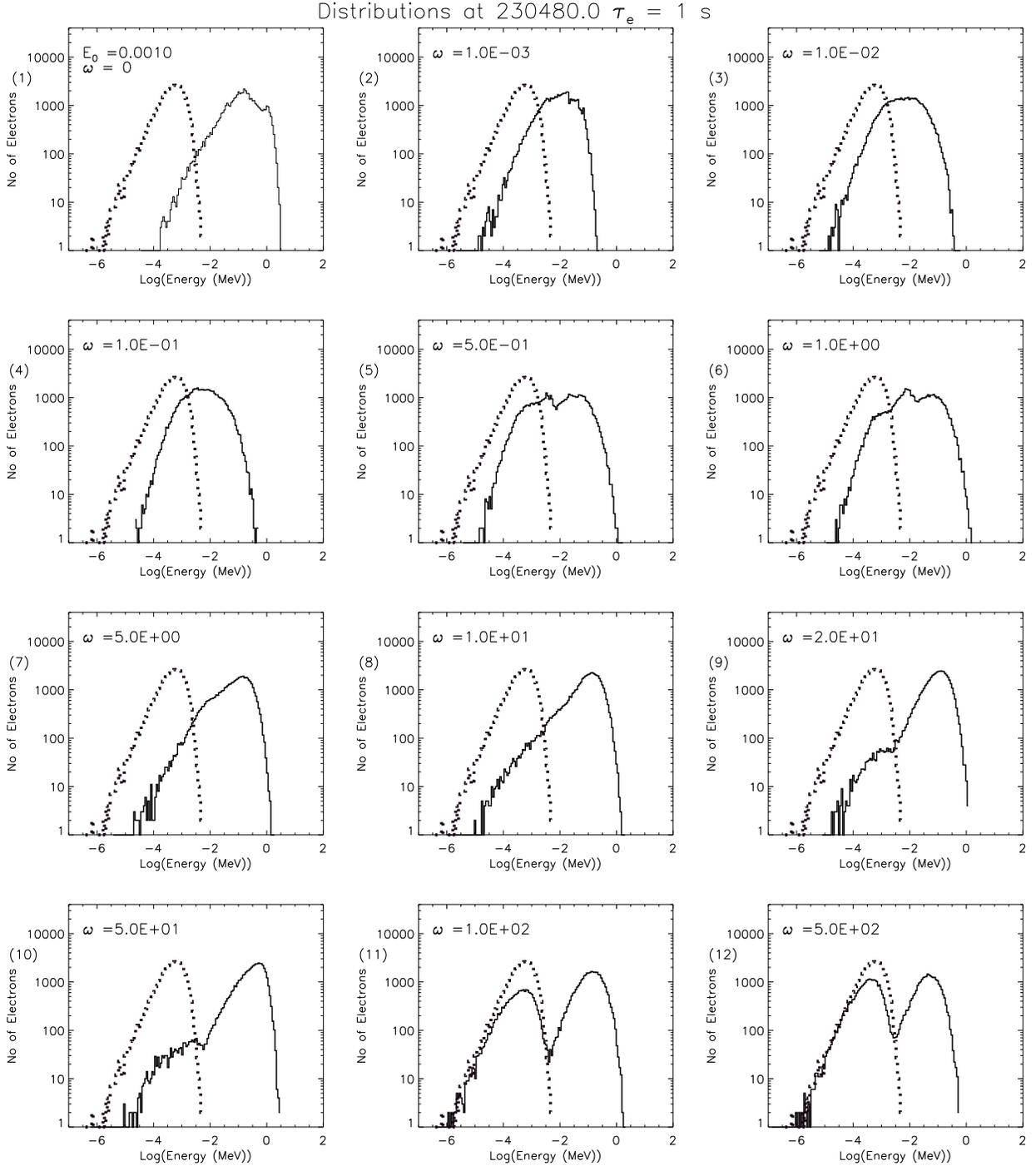}
\caption{Electron distributions for different frequencies of the 
electric field. The initial Maxwellian distribution is shown in dotted line on each panel. 
The magnitude of the electric field
is $\bar{E}_0=0.001$. The maximum integration time of each particle trajectory is 230480.
\label{fig:Ele-Energy}}
\end{figure*}

In Fig.~\ref{fig:Ele-Energy} we plot histograms of the logarithm of the initial and final 
energy distributions of electrons, for magnitude of the electric field $\bar{E}_0=0.001$ 
and total number of timesteps 230480. Again we generate each distribution using 50000 test electrons, and
show the initial Maxwellian distribution in dotted lines in each panel. 
Panel (1) shows the energy distributions for non-varying electric field ($ \omega =0$). The final energy distribution (shown in solid line)
includes a small beamlike component. The lower energy part of the final distribution is  Maxwellian-like peaking at kinetic energy
$K_{kin}=10^{-1}$ MeV, with a small beamlike component superposed at around kinetic energy $K_{kin}=1$ MeV.
The frequency of the time-varying electric field increases progressively from $ \omega =0.001$ (panel 2) to $ \omega =500$ (panel 12). The bulk of the distribution is accelerated for the frequency range $\omega =0.001$ to $\omega =50$. Accelerated distribution for $ \omega =0.001$ (panel 2) is Maxwellian-like peaking around $\sim K_{kin}= 10^{-2}$ MeV. 
From $ \omega =0.01$ (panel 3) to $ \omega =1$ (panel 6) the accelerated distributions have maximum
close to $\sim K_{kin}= 0.01$ MeV with maximum energy close to 1 MeV.
From $ \omega =5$ (panel 7) to $ \omega =20$ (panel 9) the accelerated distributions have maximum
close to $\sim K_{kin}= 0.1$ MeV with maximum energy exceeding to 1 MeV.
The final energy distributions are bi-modal for $ \omega =50 $ to $ \omega =500 $ containing a Maxwellian-like part at the energy range of the initial Maxwellian distribution and an accelerated part at higher energies with peak in the range $\sim K_{kin}= 0.03$ to $\sim K_{kin}= 0.7$ MeV. For higher frequencies the energy distributions do not show significant energy changes in the time of maximum 1 s.

\begin{figure}
\centering
   \includegraphics[width=0.45\textwidth]{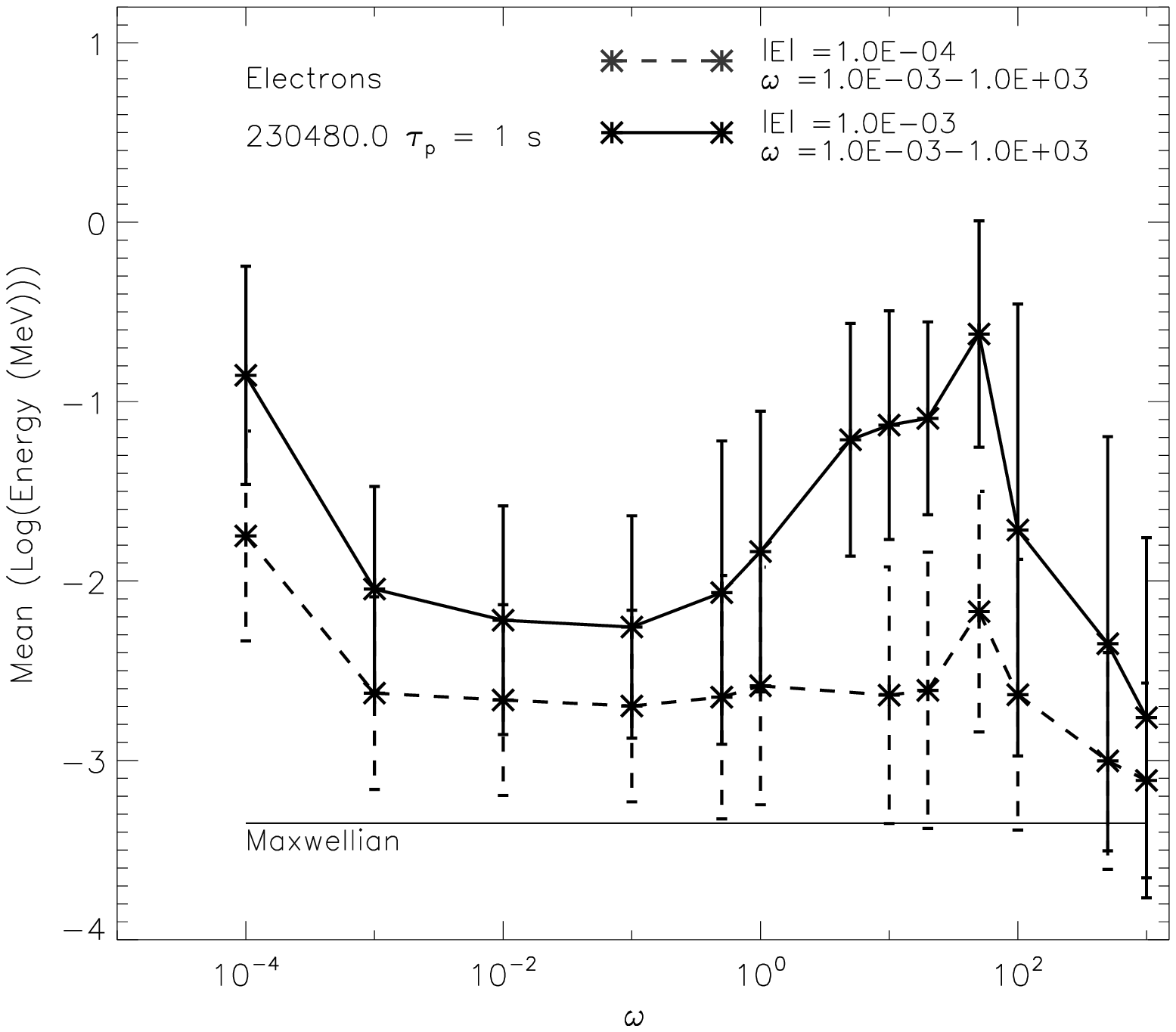}
\caption{Mean Energy and standard deviation of electron distributions for two amplitudes of the 
electric field ($\bar{E}_0 = 0.001, 0.0001$) and for range of frequencies. 
The mean energy of the initial Maxwellian distribution is shown as a straight full line. 
The constant electric
field case is represented in $\omega = 0.001$.}
\label{fig:MeanStdEne-x34}
\end{figure}

The electron distribution gains energy for most of the frequencies 
of the electric field that we used in this model except for the highest frequencies. 
In Fig.~\ref{fig:MeanStdEne-x34} we plot the mean of the logarithm of the electron energy distributions versus $\omega$ 
for two amplitudes of the electric field $\bar{E}_0 = 0.0001$ (dashed star line), $\bar{E}_0 = 0.001$ (solid star line). In the same plot we superposed as error the standard deviation in the mean for each
distribution. The mean energy for the constant
electric field is represented on this plot by $\omega= 10^{-3}$. 
The highest energy gain achieved is for electric field with $\omega= 50 $ for 
amplitude of the electric field $\bar{E}_0 = 0.001$. When $\bar{E}_0 = 0.0001$ electrons gain most energy for the constant
electric field and for $\omega= 50 $.

In the Tables that follow we summarized the energy gain aspects of the
acceleration mechanism. For each particle we
find the energy gain using its initial and final kinetic energy, that is:
\be
\frac{\Delta K^i}{K^i_{initial}} = \frac{K^i_{final} - K^i_{initial}}{K^i_{initial}} \label{eq:percent}
\ee
where $K^i_{initial}$ is the initial energy of the $i$-th particle
and $K^i_{final}$ is the final energy of the $i$-th particle. ${\Delta K^i}/{K^i_{initial}} = A$,
where $A$ takes the values 1, 10, 100. 

In Table~\ref{ta:ProtonsX3} we list the accelerated percentage of the final proton distribution for each
electric field frequency and for $E_0 = 0.001$. The first column lists the frequencies of the
electric field. The second, third and fourth columns list the percentage of the final proton distribution
for which $A$ is greater than 1, 10 and 100 respectively (Eq.~\ref{eq:percent}).
The fifth column lists the percentage of the final proton distribution that has energy greater than 1 MeV
and the last column list the highest energy in the final proton distribution in MeV.
We see that depending on the frequency of the electric field, $\sim 0.2\%$ to $\sim 17.9\%$ of the proton
distributions get accelerated to $\gamma$-ray producing energies in 1 s.  

\begin{table}
\caption{Percentage of Accelerated Protons for $E_0 = 0.001$ }             
\label{ta:ProtonsX3}      
\centering                          
\begin{tabular}{c c c c c c}        
\hline\hline                 
$\omega$ & $> 1$ & $> 10$ & $> 100$ & $>1$MeV & (MeV) \\    
\hline\hline                        
 0.0   & 89.1 \% & 50.7  \% & 14.5 \% & 3.0 \% & 15.9 \\ 
 0.001 & 59.7 \% & 12.5 \% & 2.9 \% & 0.01 \% & 1.31 \\      
  0.01 & 47.9 \% & 24.3 \% & 8.8 \% & 0.9 \% & 5.86 \\
  0.05 & 51.0 \% & 19.0 \% & 11.1 \% & 2.9 \% & 8.0 \\
   0.1 & 55.2 \% & 22.9 \% & 14.3 \% & 3.9 \% & 5.9 \\
   0.2 & 52.9 \% & 27.7 \% & 17.8 \% & 5.2 \% & 5.0 \\
   0.5 & 72.7 \% & 52.0 \% & 37.4 \% & 17.9 \% & 18.5 \\
   1.0 & 82.3 \% & 51.4 \% & 32.6 \% & 11.1 \% & 9.33 \\
  2.0 & 30.9 \% & 4.2 \% & 2.4 \% & 0.8 \% & 7.33 \\
   5.0 & 6.0  \% & 3.3 \% &  2.3 \% & 0.2 \% & 3.8\\
  10.0 & 0.3  \% & 0.1 \% &  0.01 \% & 0.0 \% & 0.31\\ 
 50.0 & 0.0 \%  & 0.0 \% &  0.0 \% & 0.0 \% & 0.005\\ 
 100.0 & 0.0 \%  & 0.0 \% &  0.0 \% & 0.0 \% & 0.005\\
\hline                                   
\end{tabular}
\end{table}
In Tables ~\ref{ta:ProtonsX4} and ~\ref{ta:ProtonsX2} we list the accelerated percentage of the final proton distribution for each available
electric field frequency for $E_0 = 0.0001$ and for $E_0 = 0.01$. The layout of these tables is the same
as for Table~\ref{ta:ProtonsX3} discussed before. 
\begin{table}
\caption{Percentage of Accelerated Protons for $E_0 = 0.0001$ }             
\label{ta:ProtonsX4}      
\centering                          
\begin{tabular}{c c c c c}        
\hline\hline                 
$\omega$ & $> 1$ & $> 10$ & $> 100$ & (MeV) \\    
\hline\hline                        
 0.0   & 26.1 \% & 4.0  \% & 1.0 \% & 0.36 \\ 
 0.001 & 9.6 \% & 1.6 \% & 0.2 \% & 0.12 \\      
  0.01 & 6.1 \% & 2.3 \% & 0.4 \% & 0.13 \\
  0.05 & 0.14 \% & 2.9 \% & 0.7 \% & 0.14 \\
   0.1 & 6.8 \% & 4.3 \% & 1.4 \% & 0.17 \\
   0.2 & 12.6 \% & 9.5 \% & 4.6 \% & 0.5 \\
   0.5 & 33.8 \% & 24.0 \% & 8.0 \% & 0.96 \\
   1.0 & 51.1 \% & 38.1 \% & 15.0 \% & 1.32 \\
  2.0 & 17.2 \% & 11.5 \% & 3.5 \% & 0.21 \\
   5.0 & 2.9  \% & 0.2 \% &  0.0 \% & 0.04\\
  10.0 & 0.01  \% & 0.0 \% &  0.0 \% & 0.012\\ 
 50.0 & 0.0 \% & 0.0 \% &  0.0 \% & 0.005\\ 
100.0 & 0.0 \% & 0.0 \% &  0.0 \% & 0.005\\ 
\hline                                   
\end{tabular}
\end{table}
\begin{table}
\caption{Percentage of Accelerated Protons for $E_0 = 0.01$ }             
\label{ta:ProtonsX2}      
\centering                          
\begin{tabular}{c c c c c c }        
\hline\hline                 
$\omega$ & $> 1$ & $> 10$ & $> 100$ & $>1$MeV & (MeV) \\    
\hline\hline                        
0.0 & 100 \% & 100 \% & 99.2 \% & 45.3 \% & 525 \\      
 0.001 & 92.9 \% & 57.1 \% & 12.8 \% & 2.6 \% & 24.5 \\      
  0.01 & 98.4 \% & 80.1 \% & 38.9 \% & 9.5 \% & 237 \\
   0.1 & 99.8 \% & 97.4 \% & 67.8 \% & 38.8 \% & 225 \\
   0.2 & 99.8 \% & 96.6 \% & 69.3 \% & 47.2 \% & 197 \\
   0.5 & 99.8 \% & 97.9 \% & 84.9 \% & 59.8 \% & 202 \\
   1.0 & 99.6 \% & 97.9 \% & 88.5 \% & 57.6 \% & 343 \\
   2.0 & 97.1 \% & 78.0 \% & 15.6 \% & 0.1 \% & 56.5 \\
  10.0 & 40.6  \% & 1.9 \% & 0.04 \% & 0.0 \% & 0.18\\ 
 50.0 & 3.0  \% & 0.03 \% & 0.0 \% & 0.0 \% & 0.005\\ 
 100.0 & 0.4 \%  & 0.0 \% &  0.0 \% & 0.0 \% & 0.005\\ 
\hline                                   
\end{tabular}
\end{table}

In Tables~\ref{ta:ElectronsX3} and ~\ref{ta:ElectronsX4} we list the accelerated percentage of the final electron distribution for each available
electric field frequency for $E_0 = 0.001$ and for $E_0 = 0.0001$ respectively.
The first, second, third and fourth columns are layed out as in Table~\ref{ta:ProtonsX3}.
The fifth column lists the percentage of the final electron distribution with energy greater than $20 \, keV$
and the last column lists the highest energy in the final electron distribution in MeV.
For $E_0 = 0.001$ and for most frequencies of the electric field (and for constant electric field) the bulk of the electron
distributions get accelerated to X-ray producing energies in the timescale of our model.  When
$E_0 = 0.0001$ only small percentage of the electron distribution accelerates to X-ray producing energies except for constant electric field where $\sim 50 \%$ accelerates to X-ray producing energies and for $\omega = 50$ where $\sim 23 \%$ accelerates to X-ray producing energies.

Flare fast electrons as revealed by hard X-ray observations generally have energy distributions characterised by
energy spectral indices in the range 2 - 5 (Dennis, \cite{Dennis85}). RHESSI data allow a less crude characterisation
of the energy distribution (Kontar et al. \cite{Kontar05}), but this range nonetheless gives a reasonable starting point for comparison with 
our results. With their high-energy peaks, sometimes bimodal in form, many of the distributions shown in Figures~\ref{fig:Pro-energy} 
and ~\ref{fig:Ele-Energy}
are clearly some way from those implied by observations. The electron distributions of Figure~\ref{fig:Ele-Energy}
would give very hard photon spectra, harder than usually observed in flares. When segments of the distributions
appeared to decline in energy in power-law form $E^{-\delta}$, we fit power-laws in energy to them, finding values 
of $\delta$ between about 1 and 2.5. Although we can account for particle acceleration in this way, to very high energies,
we have to appeal to some other agent to redistribute energy among the accelerated particles to
be compatible with observations. Very hard energy distributions are also found in most other calculations
of acceleration in either one or many dissipation regions (e.g. Turkmani et al., \cite{Turkmami06}). 

\begin{table}
\caption{Percentage of Accelerated Electrons for $E_0 = 0.001$ }             
\label{ta:ElectronsX3}      
\centering                          
\begin{tabular}{c c c c c c}        
\hline\hline                 
$\omega$ & $ > 1$ & $ > 10$ & $ > 100$ & $ >20 \,$keV & (MeV) \\    
\hline\hline                        
 0     & 99.6 \% & 97.0 \% & 78.9 \% & 91.6 \% & 3.7\\   
 0.001 & 96.7 \% & 71.9 \% & 4.8 \%  & 29 \% & 0.3\\      
 0.01  & 87.8 \% & 59.5.6 \% & 6.8 \%  & 22.8 \% & 0.6 \\
 0.1   & 92.7 \% & 50.4 \% & 10.8 \%  &  20.9 \% & 0.7 \\
 0.5   & 80.4 \% & 60.4 \% & 22.8 \%  &  39.6 \% & 1.8 \\
 1.0   & 89.3 \% & 72.2 \% & 29.5 \%  & 44.5 \% & 1.3 \\
 5.0   & 99.1 \% & 92.8 \% & 59.3 \%  & 78.6 \% & 2.1 \\
10.0   & 98.3 \% & 93.2 \% & 66.8 \%  & 85.0 \% & 1.8 \\
20.0   & 98.7 \% & 95.8 \% & 68.1 \%  & 89.6 \% & 1.5 \\ 
50.0   & 98.9 \% & 97.2 \% & 87.2 \%  & 94.9 \% & 3.2 \\
100.0  & 78.3 \% & 68.4 \% & 49.8 \%  & 65.3 \% & 2.2 \\
500.0  & 54.3 \% & 50.6 \% & 20.8 \%  & 43.9 \% & 0.6 \\
1000.0   & 41.3 \% & 33.0 \% & 3.9 \%  & 23.2 \% & 0.3 \\
10000.0   & 0.1 \% & 0.00 \% & 0.00 \%  & 0.00 \% & 0.005\\
\hline                                   
\end{tabular}
\end{table}
\begin{table}
\caption{Percentage of Accelerated Electrons for $E_0 = 0.0001$ }             
\label{ta:ElectronsX4}      
\centering                          
\begin{tabular}{c c c c c c}        
\hline\hline                 
$\omega$ & $ > 1$ & $ > 10$ & $ > 100$ & $ >20 \,$keV & (MeV) \\    
\hline \hline                       
 0     & 98.7 \% & 88.5 \% & 0.0 \% & 50.4 \% & 0.43  \\   
 0.001 & 75.5 \% & 31.6 \% & 0.0 \%  & 3.4 \% & 0.07 \\      
 0.01  & 76.4 \% & 21.7 \% & 0.6 \%  & 2.8 \% & 0.12 \\
 0.1   & 72.5 \% & 22.1 \% & 1.9 \%  &  2.0 \% & 0.097 \\
 0.5   & 60.6 \% & 31.9 \% & 4.3 \%  &  7.5 \% & 0.18 \\
 1.0   & 65.5 \% & 34.7 \% & 5.0 \%  & 9.3 \% & 0.13 \\
 10.0   & 63.0 \% & 32.4 \% & 4.0 \%  & 6.9 \% & 0.1 \\
20.0   & 61.9 \% & 38.4 \% & 4.7 \%  & 7.7 \% & 0.082 \\ 
50.0   & 84.6 \% & 62.0 \% & 11.2 \% & 23.0 \% & 0.28 \\
100.0  & 60.2 \% & 35.0 \% & 3.6 \%  & 7.2 \% & 0.11 \\
500.0  & 39.5 \% & 11.7 \% & 0.5 \%  & 0.1 \% & 0.036 \\
1000.0   & 30.3 \% & 6.7 \% & 0.2 \%  & 0.0 \% & 0.022 \\
\hline                                   
\end{tabular}
\end{table}

\begin{figure}
\centering
   \includegraphics[width=0.45\textwidth]{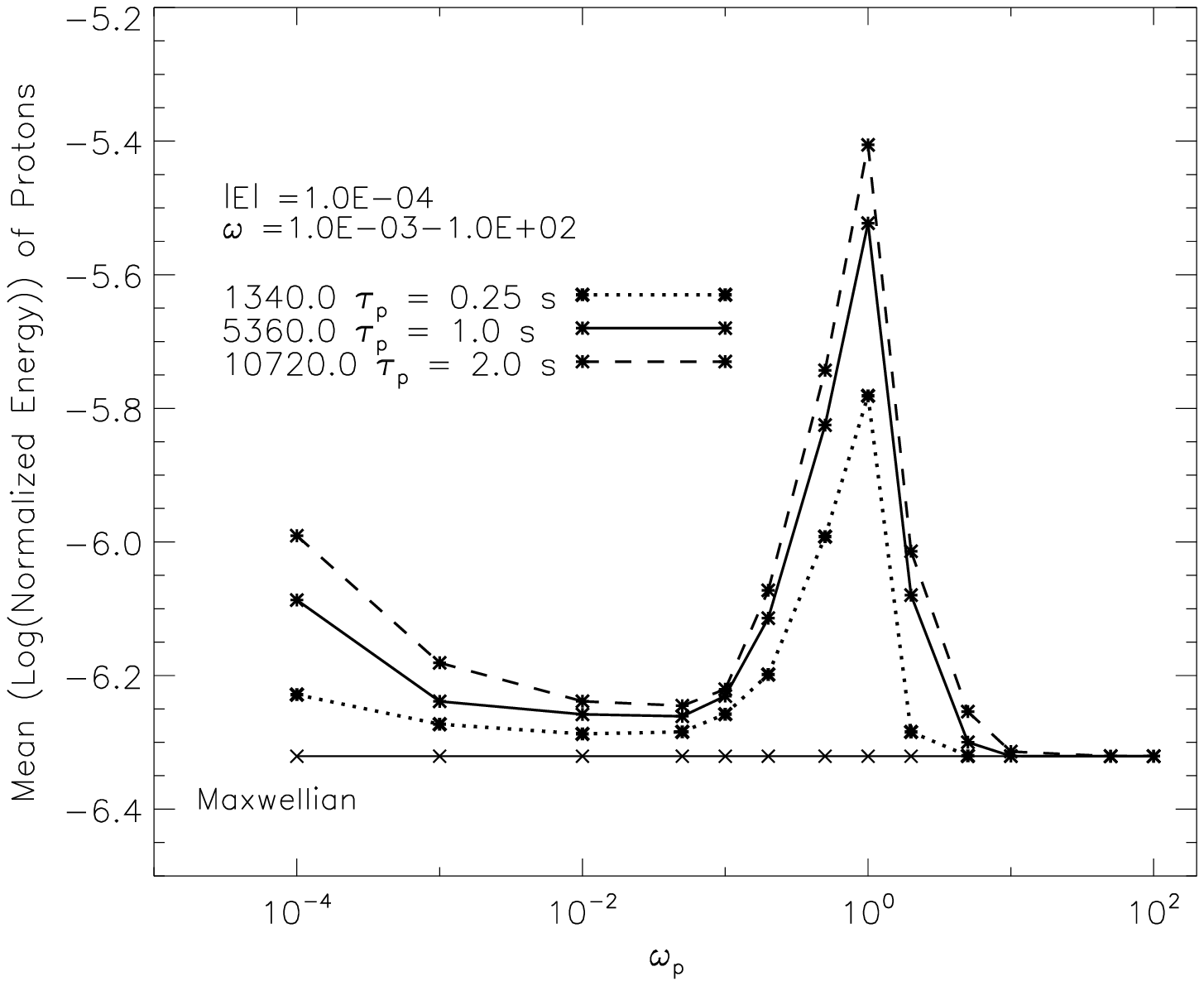}
\caption{Time evolution of logarithm of mean energy of protons for amplitude of the 
electric field ($\bar{E}_0 = 0.0001$) for constant electric field (represented on the 
graph by 0.0001) and for range of frequencies 0.001 to 100.0. 
}
\label{fig:MeanPro_x4vtime}
\end{figure}

By fixing the duration of the integrations at 1 second we generate a snapshot of the distributions produced
as particle acceleration proceeds. Obviously, particle energies will be less for shorter periods and 
greater for longer ones. As an illustration, in Figure~\ref{fig:MeanPro_x4vtime} we plot the time evolution of the logarithm of the 
mean energy of proton distributions for $\bar{E}_0 = 0.0001$, as functions of $\omega$. Mean energy is plotted for
0.25 s (1340 $\tau_p$), 1 s, and 2 s (10720 $\tau_p$). We observe that the greatest changes in mean proton energy take place for the lowest frequencies $\omega$. At high frequencies, on the other
hand, proton mean energy apparently changes little after 1 s. A sort of steady state is approached.
Since no particles escape, this indicates a decrease with proton energy of the energy increment experienced on each return to the dissipation region: protons that can be accelerated at all no longer gain much energy after this time.

\section{Discussion and conclusions}
\label{sec:discon}

In this work we investigate the particle acceleration
consequences of time-varying electric fields superposed on a X-type magnetic field to mimic generic features of dynamic, collisionless reconnection (Craig and McClymont \cite{Craig91}). We have shown that
protons and electrons may gain relativistic energies in times $\leq $1 s;
for plausible (small) electric field amplitudes and active region magnetic 
fields. This parametric study is meant to be complementary to Petkaki and 
MacKinnon (\cite{Petkaki97}) where we 
attempted to tie this test-particle approach self-consistently to an MHD description of
the passage of a wave. Although there are some qualitative similarities with the particle
behaviour studied analytically by Litvinenko (\cite{Litvinenko03}), his adoption of a spatially uniform 
electric field makes direct comparison difficult. 

Before discussing some consequences of our results, we
note some limitations of our calculation. First, this is a test
particle approach. Particles do not interact with each other, nor do
they influence the background field.  In particular, the particle
distribution including the accelerated component may well be unstable
to growth of various sorts of waves.  Obviously such wave growth would
influence the motion of particles, but we neglect this possibility. We
neglect also radiation losses.  In the solar corona this is not a
serious neglect (even for 10 MeV electrons the radiative energy loss
time is $\sim 3000$ s), but elsewhere in the cosmos it could become
significant.

We followed Craig and McClymont (\cite{Craig91}) in assuming a smoothly varying
X-type field through the whole of our system. We took the active region 
lengthscale of $10^9$ cm to define the boundary of the system. It is possible
that the field strength increases more rapidly from the neutral point,
approaching a constant value at smaller distance. This would reduce both the
adiabaticity radii of particles and the characteristic timescale.
If we continued to scale the electric field region with the proton adiabaticity radius, 
nothing would change except that the integration periods correspond to smaller real times. 
Thus particle acceleration would proceed more rapidly; however, fewer particles would be 
involved. Further consideration of this question might proceed via study of more realistic 
configurations including a neutral point or sheet (e.g. Forbes and Priest \cite{Forbes95}; 
Fletcher \& Martens \cite{Fletcher98};
Titov and D\'{e}moulin \cite{Titov99}) or complex magnetic field structures 
(e.g. Malara et al. \cite{Malara00}).

The finite width of the
nonadiabatic region allows particles to gain or lose some energy
randomly before returning to adiabatic motion. Together with repeated
encounters with the dissipation region, the consequence of mirrorings
in the extended configuration, this results in a Fermi-type,
'stochastic' acceleration.  

In our model particle acceleration
takes place for geometrical reasons. The test particle calculation is
numerically simpler than self-consistent approaches (e.g. Vlasov 
simulations, see Petkaki et al. (\cite{Petkaki03}, \cite{Petkaki06}) and gives useful
insights to the particle energization process.
There is no threshold for this type
of acceleration, unlike resonant interaction with low-frequency, MHD
waves. The necessity for protons particularly to have threshold
energies of around 25 keV is a well known difficulty when such
mechanisms are invoked (e.g. Forman et al., \cite{Forman86}). Our results indicate
that low-frequency waves may themselves perform the 'first-step'
acceleration, if they propagate in a coronal structure including a
neutral point. This may occur independently of, or simultaneously with,
the resonant cascade scenario of Miller and Vinas (\cite{Miller93}). Possible
difficulties with the number of pre-accelerated particles may be
obviated if many neutral points are present, although
such a situation obviously needs separate investigation 
(Kliem \cite{Kliem94}).

Most of the resulting proton distributions have a bi-modal form
(see Fig.~\ref{fig:Pro-energy}). Electron distributions are also bi-modal for the highest frequencies, $20 \leq \omega \leq 500$  (see Fig.~\ref{fig:Ele-Energy}). Whereas for the lowest frequencies of the electric field the bulk of the initial electron Maxwellian distribution is 
accelerated, for the highest frequencies only part of the electron distribution is accelerated (Table~\ref{ta:ElectronsX3}). Acceleration occurs for all frequencies $\omega \leq 10$ when addressing the proton distributions (Table~\ref{ta:ProtonsX3}).
The bi-modal form of the proton energy distributions might offer a way to have protons of gamma-ray producing energies ($K_{kin} \sim $ 2 MeV) without 
the energetically dominant population at lower energies that is the inevitable
consequence of a diffusive particle accelerator (see the Appendix of
Eichler (\cite{Eichler}); and MacKinnon (\cite{MacKinnon91})). Investigation of the velocity space stability 
of these distributions needs details of the angular distribution at particular points in space,
and is not discussed here.

We note the effectiveness of acceleration of the two species varies
according to the frequency of oscillation invoked. Electrons are
accelerated for a broader spectrum of frequencies. Frequencies 0.001 to 1000 
have been simulated here, corresponding to real frequencies in the range 5 Hz to 5 MHz
(cf. the frequency range of waves from the base of the solar corona, probably in the range 0.01 Hz to
10KHz, e.g. Marsch et al. (\cite{Marsch82})). 
Frequencies lower than 0.001 will also accelerate electrons as indicated by the
net acceleration achieved for the constant electric field cases (Fig.~\ref{fig:MeanStdEne-x34}),
but frequencies higher than 1000 do not produce a net acceleration in the timescale of our model.

Considered as a function of $\omega$, the mean energy of the accelerated electron distribution exhibits a peak
in the broad range $5 < \omega < 100$. Such a peak leads us to suspect a resonance involving two or 
more of the timescales in the problem. The initial gyrofrequencies of electrons lying in the adiabatic portion of the
dissipation region also generally lie in this range. Inverse crossing times ($1/t_{cr}$, see Equation~\ref{eq:cross})
comparable with $\omega$ might also lead to enhanced acceleration. Using Equation~\ref{eq:cross}, but taking account
also of the mean increase in $u_{x,y}$ we do indeed find upper limits in the range $5 < 1/t_{cr} < 100$. 

Protons are accelerated for low electric field frequencies, achieving $\gamma$-ray producing
energies in 5360 $\tau_p = 1$ s for frequencies
$\omega < 10$ and for $E_0 = 0.001$ and $E_0 = 0.01$. A local peak in the mean energy of the accelerated proton distribution is seen at
$0.1 < \omega < 2.0$. This range of frequencies are comparable to the gyrofrequencies of protons in the adiabatic region 
for our set of initial conditions and to the proton inverse crossing time. 

The variability of the effectiveness of acceleration of the two species according to the frequency of electric field oscillation might bear on the apparent
variation of electron/proton ratios in flares (Ramaty \& Murphy \cite{Ramaty87}) and the
phenomenon of `electron-only' flares (Rieger \cite{Rieger89}). As a general comment, we note that higher
frequency disturbances favour electrons over ions, although more definitive
statements will need a proper treatment involving a more realistic
wave.

For most frequencies and for constant electric field, part of the electron
distribution escapes from the system boundaries before 230400$\tau_e = 1$ s. 
Electrons on average escape in less than 0.6 seconds in the frequency range
$10 \le \omega \le 50$.
Protons on the other hand do not escape the system boundaries on the same
timescale of 5360$\tau_p = 1$ s for $E_0 = 0.001$ and $E_0 = 0.0001$.
Electrons are accelerated
more rapidly than protons to energies that do not allow them to mirror inside our system
boundaries.

Here and in Petkaki \& MacKinnon (\cite{Petkaki97}) we investigate particle acceleration 
at a null in the presence of a linear disturbance. Such calculations may give some insight into particle acceleration in
flares, although conditions then presumably depart severely from linearity, but might be most relevant to 
quiescent, long-lasting phenomena such as radio noise storms. Definitely involving deka-keV electrons (Raulin \& Klein, \cite{Raulin94}) and showing correlations with X-ray variations, but without 
chromospheric, flare-like signatures
(Svestka et al \cite{Svestka82}; Crosby et al. \cite{Crosby96}), particle acceleration in noise storms 
might occur as described here, if the relevant coronal structures include null points. 
Electrons accelerated at a neutral point will likely encounter very large mirror ratios, trapping them
in the corona (Fletcher \& Martens 
\cite{Fletcher98}) and accounting for the 
exclusively coronal phenomena accompanying noise storms.

\end{document}